\documentclass[aps,prl,twocolumn,showpacs,preprintnumbers,amsmath,amssymb,superscriptaddress]{revtex4}%

\usepackage{graphicx}%
\usepackage{dcolumn}
\usepackage{amsmath}
\usepackage{color}

\begin{document}

\title{Pressure Induced Static Magnetic Order in Superconducting FeSe$_{1-x}$}
\author{M.~Bendele}
 \affiliation{Laboratory for Muon Spin Spectroscopy, Paul Scherrer
Institut, CH-5232 Villigen PSI, Switzerland}
 \affiliation{Physik-Institut der Universit\"{a}t Z\"{u}rich,
Winterthurerstrasse 190, CH-8057 Z\"urich, Switzerland}
\author{A.~Amato}
 \affiliation{Laboratory for Muon Spin Spectroscopy, Paul Scherrer
Institut, CH-5232 Villigen PSI, Switzerland}
\author{K.~Conder}
 \affiliation{Laboratory for Developments and Methods, Paul Scherrer Institute,
CH-5232 Villigen PSI, Switzerland}
\author{M.~Elender}
 \affiliation{Laboratory for Muon Spin Spectroscopy, Paul Scherrer
Institut, CH-5232 Villigen PSI, Switzerland}
\author{H.~Keller}
 \affiliation{Physik-Institut der Universit\"{a}t Z\"{u}rich,
Winterthurerstrasse 190, CH-8057 Z\"urich, Switzerland}
\author{H.-H.~Klauss}
 \affiliation{IFP, TU Dresden, D-01069 Dresden, Germany}
\author{H.~Luetkens}
 \affiliation{Laboratory for Muon Spin Spectroscopy, Paul Scherrer
Institut, CH-5232 Villigen PSI, Switzerland}
\author{E.~Pomjakushina}
 \affiliation{Laboratory for Developments and Methods, Paul Scherrer Institute,
CH-5232 Villigen PSI, Switzerland}
\author{A.~Raselli}
 \affiliation{Laboratory for Developments and Methods, Paul Scherrer
Institut, CH-5232 Villigen PSI, Switzerland}
\author{R.~Khasanov}
 \email[Corresponding author: ]{rustem.khasanov@psi.ch}
 \affiliation{Laboratory for Muon Spin Spectroscopy, Paul Scherrer
Institut, CH-5232 Villigen PSI, Switzerland}

\begin{abstract}
We report on a detailed investigation of the electronic phase
diagram of FeSe$_{1-x}$ under pressures up to 1.4~GPa by means of
AC magnetization and muon-spin rotation. At a pressure
$\simeq0.8$~GPa the non-magnetic and superconducting FeSe$_{1-x}$
enters a region where long range static magnetic order is realized
above $T_c$ and bulk superconductivity coexists and competes on
short length scales with the magnetic order below $T_c$. For even
higher pressures an enhancement of both the magnetic and the
superconducting transition temperatures as well as of the
corresponding order parameters is observed. These exceptional
properties make FeSe$_{1-x}$ to be one of the most interesting
superconducting systems investigated extensively at present.
\end{abstract}

\maketitle

%

The phase diagram of the recently discovered Fe-based
high-temperature superconductors (HTS) \cite{Kamihara08} share a
common feature with cuprates and heavy fermions. The parent
compounds  of the Fe-based HTS, such as, LnOFeAs (Ln=La, Ce, Pr,
Sm) \cite{Cruz08,Klauss08,Luetkens09, Zhao08,Zhao08_2,
Carlo09,Drew09,Sanna09}, AFe$_2$As$_2$ (A = Ba, Sr, Ca)
\cite{Jesche08,Huang08,Zhao08_3,Goko09} and Fe(SeTe)
\cite{Li09,Khasanov09_FeSeTe} exhibit long-range static magnetic
order. Upon doping or application of pressure (chemical or
external), magnetism is suppressed and superconductivity emerges.
Recent investigations revealed, however, that the structurally
most simple binary compound FeSe$_{1-x}$ is an exception of this
rule \cite{Medvedev09}. Different from the other Fe-based HTS,
FeSe$_{1-x}$ did not seem to exhibit static magnetic order for
pressures up to about 30~GPa \cite{Medvedev09}. Yet, short-range
spin fluctuations, which are strongly enhanced towards $T_c$, were
observed \cite{Imai09}.
The superconducting transition temperature of FeSe$_{1-x}$ was
found to increase continuously to $\simeq 37$~K at $7-9$~GPa. For
higher pressures  a decrease is observed with $T_c\simeq6$~K
approaching $20$~GPa \cite{Medvedev09,Margadonna09}. Subsequent
experiments with finer pressure steps revealed, however, a local
minimum on $T_c(p)$ at $1.5$~GPa of unexplained nature
\cite{Miyoshi09}.

In this letter we report on a detailed study of the evolution of
the superconducting and magnetic properties of FeSe$_{1-x}$  as a
function of pressure and temperature through a combination of AC
susceptibility and muon-spin rotation ($\mu$SR) techniques. Two
samples with the nominal composition FeSe$_{0.94}$ and
FeSe$_{0.98}$ were investigated. The obtained phase diagram of
FeSe$_{1-x}$ was found to be separated into three distinct
regions.
At low pressures, $0\leq p\lesssim 0.8$~GPa, the samples are
nonmagnetic and $T_c$ increases monotonically with increasing
pressure.
In the intermediate pressure region, $0.8\leq
p\lesssim1.0$~GPa, $T_c(p)$ decreases with increasing pressure and
static magnetism develops. In this region of the phase diagram
the superconducting and the magnetic order parameters coexist and
compete on a short length scale. Incommensurate magnetic order,
which sets in above $T_c$, becomes partially (or even fully)
suppressed below $T_c(p)$.
At higher pressures, $p\gtrsim1.0$~GPa, $T_c(p)$ shows a tendency
to  rise again. The magnetic order becomes commensurate and both,
bulk magnetism and bulk superconductivity coexist within the
whole sample volume.

FeSe$_{1-x}$ samples with the nominal composition FeSe$_{0.94}$
and FeSe$_{0.98}$ were prepared by solid state reaction similar to
that described in
Refs.~\onlinecite{Hsu08,MargadonnaChemComm2008,Pomjakushina09}.
Powders of minimum purity 99.99\% were mixed in appropriate
ratios, pressed and sealed in a double-walled quartz ampoule. The
pressed rod was heated up to 700$^{\rm o}$C followed by annealing
at 400$^{\rm o}$C \;\cite{Pomjakushina09}.

The pressure was generated in a piston-cylinder type of cell
especially designed to perform muon-spin rotation experiments
under pressure \cite{Andreica01}. As a pressure transmitting
medium 7373 Daphne oil was used. The pressure was measured in situ
by monitoring the pressure shift of the superconducting transition
temperature of Pb or/and In. Two types of cells, the first one made
from CuBe alloy [maximum pressures
$p_{max}(300{\rm~K})\simeq1.4$~GPa and
$p_{max}(7{\rm~K})\simeq1.1$~GPa] and the second one made from
MP35 alloy [$p_{max}(300{\rm~K})\simeq1.7$~GPa and
$p_{max}(7{\rm~K})\simeq1.4$~GPa], were used.

AC susceptibility measurements were performed by using a home made
AC magnetometer with a measuring field $\mu_0H_{\rm AC}\sim0.1$~mT
and frequency $\nu=96$~Hz. In order to keep the position of the
sample unchanged during the series of AC susceptibility under
pressure measurements, the excitation and the two pick-up coils
were wound directly on the cell. To ensure that the AC signal was
entirely determined by the Meissner response of individual grains
and not by the Josephson type of weak links between them,
measurements of the AC susceptibility as a function of $\nu$
($0\leq\nu\leq599$~Hz) and $H_{AC}$ ($0\leq
\mu_0H_{AC}\leq0.5$~mT) at $T=2.5$~K on a standard Quantum Design
PPMS instrument were performed. The experiments reveal that the AC
magnetization ($M_{AC}$) scales linearly with $H_{AC}$ and is
independent on $\nu$ as expected for a superconductor in the
Meissner state.

The zero-field muon-spin rotation (ZF $\mu$SR) experiments were
carried out at the $\mu$E1 beam line at the Paul Scherrer
Institute, Switzerland for the temperatures ranging  from 0.25 to
50~K. The typical counting statistics were $\sim7\cdot 10^{6}$
positron events for each particular data point.

\begin{figure}[htb]
\includegraphics[width=1.0\linewidth]{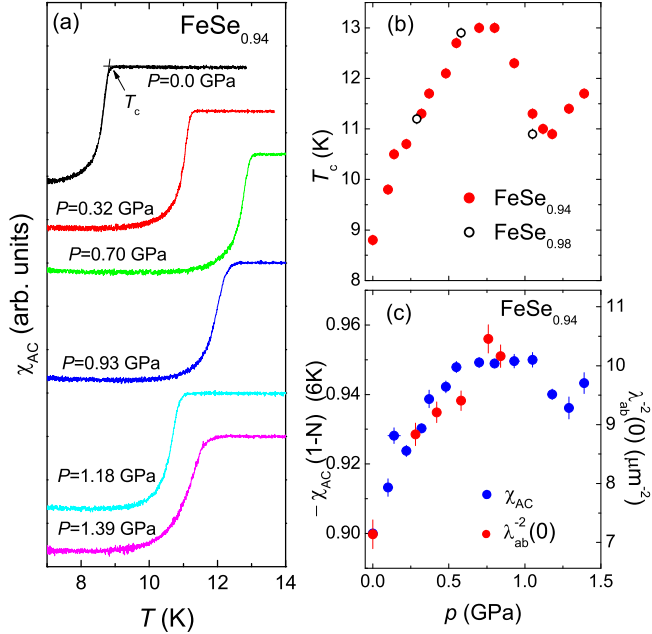}
%
\caption{(a) Temperature dependence of $\chi_{AC}$ of
FeSe$_{0.94}$ measured at (from the top to the bottom) $p=0.0$,
0.32, 0.7, 0.93, 1.18, and 1.39~GPa. The transition temperature
$T_{\rm c}$ is determined from the intersection of straight line
fits to the data above and below the transition. (b) Dependence of
$T_c$ on $p$ of FeSe$_{0.94}$ and FeSe$_{0.98}$. (c) Pressure
dependence of the normalized AC susceptibility $-\chi_{\rm
AC}\;(1-N)$ at $T=6$~K and the inverse squared in-plane magnetic
penetration depth $\lambda_{ab}^{-2}$ at $T=0$~K
\cite{Khasanov08_FeSe_pressure_lambda}. }
 \label{fig:AC}
\end{figure}

The response of the superconducting state of FeSe$_{1-x}$ to
pressure was studied in AC susceptibility ($\chi_{AC}$)
experiments, Figure~\ref{fig:AC}. The transition temperature
$T_c$, Figure~\ref{fig:AC}b, shows a non-monotonic increase with
a local minimum at $p\simeq1.2$~GPa, similar to $T_c(p)$ reported
by Miyoshi {\it et al.} \cite{Miyoshi09}.
Figure~\ref{fig:AC}c depicts $-\chi_{\rm AC}(T)\; (1-N)$ at
$T=6$~K. Here $N$ denotes the demagnetization factor which,
assuming a spherical shape of the sample grains, was taken to be
equal to 1/3. For temperatures lower than $T_c$, $|\chi_{\rm
AC}(T)\;(1-N)|$ is smaller than unity due to the penetration of
the AC magnetic field on a distance $\lambda$ from the surface of
each individual grain ($\lambda$ is the magnetic penetration
depth). Following Shoenberg \cite{Shoenberg54}, $\chi_{\rm AC}$ in
a granular sample is expected to scale with $\lambda^{-2}$, which
is the case here, as may be seen from the comparison of $\chi_{\rm
AC}(6$~K) with $\lambda_{ab}^{-2}(T=0)$ obtained in transverse
field $\mu$SR experiment \cite{Khasanov08_FeSe_pressure_lambda}.

\begin{figure}[htb]
\includegraphics[width=1.0\linewidth]{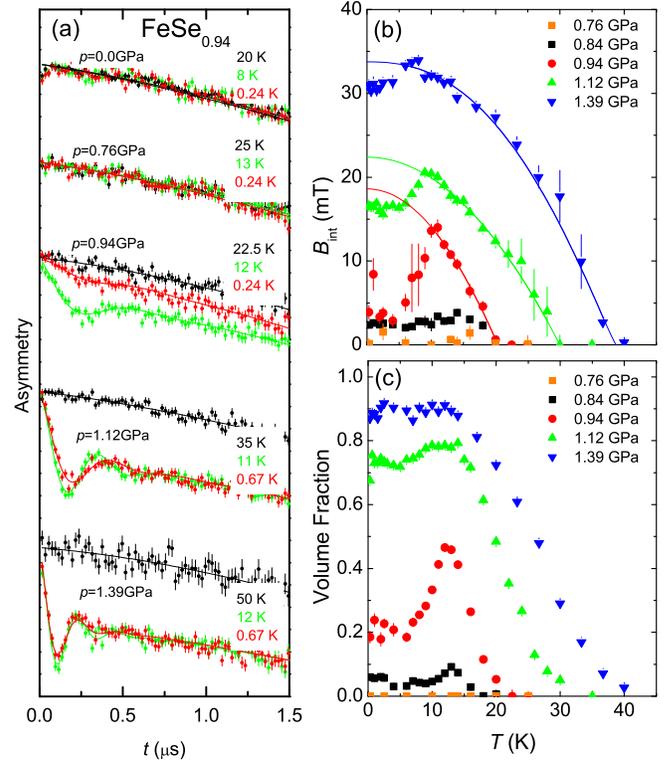}
%
\caption{(a) Zero-field $\mu$SR time-spectra of
FeSe$_{0.94}$ me\-asu\-red  at (from the top to the bottom)
$p=0.0$, 0.76, 0.94, 1.12, and 1.39~GPa.  Dependence of the
internal field at the muon stopping site $B_{int}$ which is
proportional to the magnetic order parameter (b), and the
magnetic volume fraction (c), on temperature at various
pressures. The solid lines in panel {\bf b} are the fit of
$B_{int}(T)$ in the region $T_c(p)\leq T\leq T_N$ to
$B_{int}(T)=B_{int}(0)[1-(T/T_N)^\alpha]^\beta$ ($\alpha$ and
$\beta$ are the power exponents ). }
 \label{fig:ZF}
\end{figure}

The magnetic response of FeSe$_{1-x}$ was studied in ZF $\mu$SR experiments.
In the following we discuss the ZF
$\mu$SR data for the three different pressure regions.

In the low-pressure region, $0\leq p\lesssim0.8$~GPa, where $T_c$
increases with increasing $p$, the ZF $\mu$SR time-spectra prove
the absence of long range magnetic order for all temperatures as
exemplified by the identical weakly damped spectra for $T=0.24$~K,
near $T_c(p)$ and 20~K, see $p=0.0$ and $p=0.76$~GPa data in
Figure~\ref{fig:ZF}a. The solid lines in Figure~\ref{fig:ZF}a
correspond to a two-component fit:
\begin{equation}
A^{ZF}(t)=A^{ZF}_{S}(t)+A^{ZF}_{PS}(t).
\end{equation}
with the first component describing the sample response and the
second one representing the contribution of the pressure cell (ZF
responses of the CuBe and MP35 cells are described in
\cite{Andreica01}).  The sample contribution is well fitted to the
single-exponential decay function \cite{Khasanov08_FeSe}:
\begin{equation}
A_S^{ZF}(t)=A_{S,0}^{ZF}\; e^{-\Lambda_0 t},
 \label{eq:ZF-LF}
\end{equation}
($\Lambda_0$ is the exponential relaxation rate), thus revealing
that very diluted and randomly oriented magnetic moments exist in
the sample volume which can be attributed to small traces of Fe
impurities, see Ref.~\onlinecite{Khasanov08_FeSe}.

In the intermediate pressure region, $0.8\leq p\lesssim1.0$~GPa,
the spontaneous muon-spin precession is clearly observed in the ZF
$\mu$SR time spectra, see $p=0.94$~GPa data in
Figure~\ref{fig:ZF}a. Therefore, long range magnetic order is
established below the N\'eel temperature $T_N$. The analysis of
the $\mu$SR data was made by accounting for the separation of the
sample into magnetically ordered regions with muons experiencing a
static local field and nonmagnetic (paramagnetic) regions:
\begin{eqnarray}
A_S^{ZF}(t)&= &A_{S,0}^{ZF}\; \left[{\rm m}\left(\frac{2}{3}\;
j_0(\gamma_\mu
B_{int}t)e^{-\Lambda_T t}+\frac{1}{3}\;e^{-\Lambda_L t}\right)\right. \nonumber \\
 &&   +(1-\textrm{m})\;e^{-\Lambda_0 t} \biggr].
\end{eqnarray}
Here $m$ is the magnetic volume fraction of the sample, $j_0$ is a
zeroth order Bessel function, $\gamma=2\pi\;135.5$~MHz/T is the
muon gyromagnetic ratio, and $\Lambda_T$ and $\Lambda_L$ are the
exponential relaxation rates longitudinal and transverse to the
initial muon-spin polarization.
The oscillating part of the signal was found
to be good described by a Bessel function, which is archetypical
for incommensurate magnetic order \cite{Savici02}. The dependence of the internal field
$B_{int}$, corresponding to the magnetic order parameter, and
the magnetic volume fraction on temperature are shown in
Figures~\ref{fig:ZF}b and c.

For pressures above  1.0~GPa we observe a further increase of
the magnetic volume fraction and of the internal magnetic field
$B_{int}$, Figure~\ref{fig:ZF}. Additionally, we find that the
$\mu$SR lineshape is better described by a damped cosine with zero
initial phase rather than by a Bessel function:
\begin{eqnarray}
A_S^{ZF}(t)&= &A_{S,0}^{ZF}\; \left[{\rm m}\left(\frac{2}{3}\;
\cos(\gamma_\mu
B_{int}t)e^{-\Lambda_T t}+\frac{1}{3}\;e^{-\Lambda_L t}\right)\right. \nonumber \\
 &&   +(1-\textrm{m})\;e^{-\Lambda_0 t} \biggr].
\end{eqnarray}
This suggests that in the high pressure region the magnetic order becomes commensurate.

\begin{figure}[htb]
\includegraphics[width=0.9\linewidth]{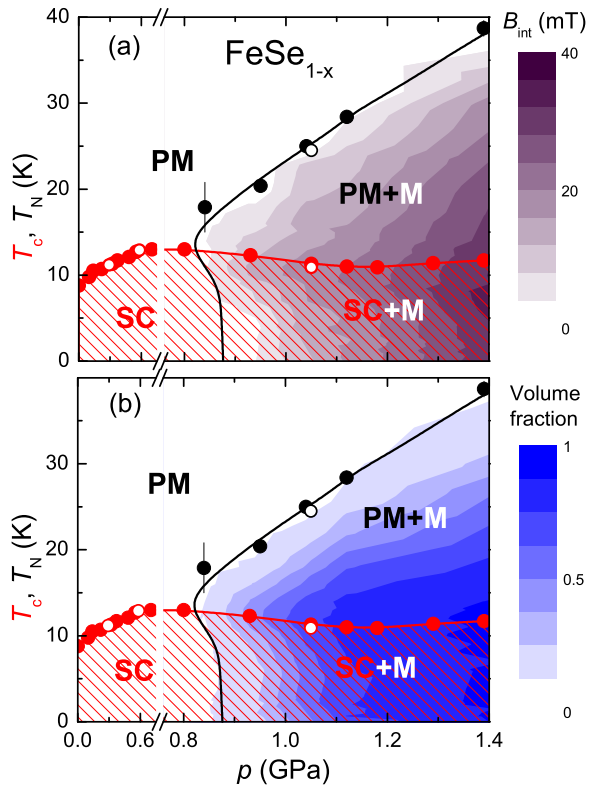}
 \vspace{-0.3cm}
\caption{ (a) Pressure dependence of the superconducting
transition temperature $T_c$, the magnetic ordering temperature
$T_N$, and the internal field $B_{int}$ (magnetic order parameter)
obtained in AC susceptibility and muon-spin rotation experiments.
(b) Pressure dependence of $T_N$, $T_c$, and the magnetic volume
fraction. The $T_c(p)$ and $T_N(p)$ lines are guides for the eye.
The closed and the open symbols refer to FeSe$_{0.94}$ and
FeSe$_{0.98}$ sample. SC, M, and PM denote the superconducting,
magnetic and nonmagnetic (paramegnetic) states of the sample.}
 \label{fig:Phase-diagramm}
\end{figure}

Figure~\ref{fig:Phase-diagramm} summarizes our results on the
magnetism and superconductivity in an electronic phase diagram for
FeSe$_{1-x}$.The magnetic order coexists and competes with
superconductivity for $p\gtrsim0.8$~GPa. Above this pressure long
range magnetic order is established below $T_N>T_c$ and bulk
superconductivity sets in below $T_c$. The competition of the two
ground states in this pressure range is evident from the following
two observations: First, as a function of pressure $T_c$ is
weakened as soon as magnetic order appears, leading to the local
maximum at $p\simeq0.8$~GPa in $T_c(p)$. Second, as
a function of temperature $B_{int}$, as well as the magnetic volume
fraction, decrease below $T_c$ showing that the magnetism, which
develops at higher temperatures, becomes partially (or even
fully) suppressed by the onset of superconductivity.
The superconducting volume fraction is close
to 100\% for all pressures, Figure~\ref{fig:AC}c, while the
magnetic fraction increases continuously and reaches $\simeq90$\%
at the highest pressure investigated $p\simeq1.39$~GPa,
Figure~\ref{fig:ZF}c. In other words, both ground states coexist
in the full sample volume at $p=1.39$~GPa. Our data do not provide
any indication for macroscopic phase separation into
superconducting and magnetic clusters (bigger than a few nm in
size), as observed e.g.\ for Ba$_{1-x}$K$_x$Fe$_2$As$_2$
\cite{Park09}. Actually, the data rather point to a coexistence of
both order parameters on an {\it atomic} scale. This scenario
is compatible with the itinerant two-band models of Fe-based
HTS proposed recently by Vorontsov {\it et al.} \cite{Vorontsov09}
and Cvetkovic and Tesanovic \cite{Cvetkovic09}. According to these
models the transition between the magnetic and the superconducting
states may involve the formation of the intermediate phase, where
both superconductivity and magnetism coexist.

In conclusion, the magnetic and superconducting properties of
FeSe$_{1-x}$ were studied as a function of pressure up to 1.4~GPa
by means of AC magnetization and muon-spin rotation techniques.
Above $\simeq0.8$~GPa superconductivity was found to coexists with
magnetism with N\'eel temperatures $T_N>T_c$. In a narrow pressure
range, where a local minimum in $T_c(p)$ is observed,
superconductivity competes with magnetism in the sense that the
magnetic volume fraction and the magnetic order parameter are
suppressed below $T_c$. At the highest pressure investigated here
superconductivity and static long range commensurate magnetism
coexist on short length scales in the full sample volume.
Furthermore, both forms of order seem to be stabilized by
pressure, since $T_c$ as well as $T_N$ and the magnetic order
parameter simultaneously increase with increasing pressure. This
exceptional observation provides a new challenge for theories
describing the mechanism of high temperature superconductivity.

This work was performed at the S$\mu$S Paul
Scherrer Institute (PSI, Switzerland).  The work of MB was supported by
the Swiss National Science Foundation. The work of EP was
supported by the NCCR program MaNEP. RK acknowledges the discussion with Z.~Tesanovic.

\end{document}